\documentclass[a4paper,11pt]{article}
\usepackage{pos}

\def\beq{\begin{equation}}
\def\eeq{\end{equation}}
\def\bea{\begin{eqnarray}}
\def\eea{\end{eqnarray}}
\def\D0{D\O }

\title{Semileptonic $B$ decays and $|V_{xb}|$ update}

\author*[a,b]{Giulia Ricciardi}

\affiliation[a]{Dipartimento di Fisica E. Pancini, Universit\`a  di Napoli Federico II \\
  Complesso Universitario di Monte Sant'Angelo, Via Cintia,
I-80126 Napoli, Italy}
\affiliation[b]{INFN, Sezione di Napoli,
  Complesso Universitario di Monte Sant'Angelo, Via Cintia,
I-80126 Napoli, Italy}

\emailAdd{giulia.ricciardi@na.infn.it}

\abstract{We review the status  of semileptonic $B$ decays and  $|V_{xb}|$ determinations.}

\FullConference{%
  BEAUTY2020\\
  21-24 September 2020\\
  Kashiwa, Japan (online)}

\begin{document}
\maketitle

\section{Introduction}
\label{intro}

Semileptonic $B$ decays are the processes of election when it comes to a precise determination of the parameters $|V_{xb}|$  of the Cabibbo-Kobayashi-Maskawa (CKM) matrix. At the lowest order in the SM, they are mediated by  tree level quark decays, and the presence of leptons in the final states simplifies the QCD analyses, since hadronic and leptonic currents factorize.
Precision studies of semileptonic $B$ meson decays have been made possible by  the large samples of  $B$ mesons collected at the $B$-factories and at LHCb, and by a concomitant progress in theoretical computations, especially in lattice QCD.
Since at least four decades, the parameters  $|V_{xb}|$  
play a central role in the analyses of the unitarity triangle and in testing the Standard Model (SM). 
A long-standing tension among their values, depending on whether they  are  extracted  using exclusive or inclusive semi-leptonic $B$ decays, has been an additional  motivation for  more and more refined theoretical and experimental analyses. 
We briefly review  recent progress on the study of semileptonic $B$ decays and on  $|V_{xb}|$ determinations~\footnote{For brief overviews see for example \cite{Ricciardi:2019zph,Ricciardi:2017lne,Koppenburg:2017mad,Ricciardi:2016pmh,Ricciardi:2014aya,Ricciardi:2014iga} and references therein.}. 

\section{Inclusive decays $B \to X_c \, \ell \nu_\ell$}

In  inclusive $ B \rightarrow X_c \, \ell \, \nu_\ell$ decays,  the final state
$X_c$ is an hadronic state originated by the charm  quark. 
The large hierarchy between the intrinsic large 'dynamic' scale of energy release and  the hadronic soft scale  leads naturally to $\Lambda_{QCD}/m_b$ as an expansion parameter of non-perturbative effects. The expansion for the total semileptonic width takes the form 
\beq
\label{HQE}
\Gamma(B\rightarrow X_c l \nu) =\frac{G_F^2m_b^5}{192 \pi^3}
|V_{cb}|^2 \left[ c_3 \langle O_3 \rangle + 
  c_5\frac{ \langle O_5 \rangle }{m_b^2}+c_6\frac{ \langle O_6 \rangle }{m_b^3}+O\left(\frac{\Lambda^4_{QCD}}{m_b^4},\; \frac{\Lambda^5_{QCD}}{m_b^3\, m_c^2}, \dots \right) \right] 
\eeq
Here  $c_d$ ($d=3,5,6 \dots$) are short distance coefficients, calculable  in perturbation theory as a series in the strong coupling $\alpha_s$, and
$O_d$ denote local operators of (scale) dimension $d$. 
The hadronic expectation values of the operators are the (normalized) forward matrix elements,  
encode the nonperturbative corrections and can be parameterized in terms of  heavy quark expansion (HQE)  parameters, whose number grows with powers of $\Lambda_{QCD}/m_b$. Similar expansions hold for sufficiently inclusive quantities as the moments of distributions of
charged-lepton energy, hadronic invariant mass and hadronic energy.

At order $1/m_b^0$ in the HQE, the  perturbative corrections up to order $\alpha_s^2$ to the width and to the moments of the lepton energy and hadronic mass
distributions are known completely \cite{Melnikov:2008qs, Gambino:2011cq, Trott:2004xc, Aquila:2005hq, Pak:2008qt, Pak:2008cp, Biswas:2009rb}. The terms of order $\alpha_s^{n+1} \beta_0^n$, where $\beta_0$ is the first coefficient of the QCD $\beta$ function, $\beta_0= (33-2 n_f)/3$, have also been computed following  the
 Brodsky-Lepage-Mackenzie procedure \cite{Aquila:2005hq, Benson:2003kp}.
 Perturbative corrections to the coefficients   of the kinetic operator  \cite{Becher:2007tk,Alberti:2012dn}
and  the chromomagnetic operator
\cite{Alberti:2013kxa, Mannel:2014xza, Mannel:2015jka}   have been
 evaluated   at order $\alpha_s$.
Two independent parameters, $\rho^3_{D,LS}$, are also needed to describe matrix elements of operators of dimension six, that is 
at order $1/m_b^3$. Their coefficients
have long been known at tree level~\cite{Gremm:1996df}, and more recently   $\alpha_s$ corrections to the coefficient of the $\rho^3_{D}$ term have been computed~\cite{Mannel:2019qel}.
Starting at  order  $\Lambda_{QCD}^3/m_b^3$,   terms with an infrared sensitivity to the charm mass appear, at this order as a $\log m_c$ contribution \cite{Bigi:2005bh,Breidenbach:2008ua, Bigi:2009ym}.
Presently, the matrix elements have been  identified and estimated up to the order  $1/m_b^{4}$ and $1/m_b^{5}$ ~\cite{Dassinger:2006md, Mannel:2010wj,Heinonen:2014dxa}.

 \section{Inclusive $|V_{cb}|$ determination}
 
 The shapes of the kinematic distributions in the $B\to X_c\ell\nu$ decays are sensitive to 
the masses of the $b$ and $c$ quarks and the non-perturbative HQE parameters, and all these  quantities are affected by the
 particular theoretical scheme used to define the quark masses.
 Non perturbative parameters can be extracted together with $|V_{cb}|$  in a global fit  based on experimentally measured distributions and momenta.
Global fit analyses  differ by the data sets they are based onto, the theoretical  scheme employed, and the order of truncation of the HQE expansion.
Challenges are experimental selections applied to the data as well as to properly account for correlations.

A recent global analysis of the inclusive $B\to X_c\ell\nu$ has been done by HFLAV
\cite{Amhis:2019ckw}. 
In the framework of kinetic scheme, $|V_{cb}|$ is extracted together with the $b$ and $c$ quark masses and 4 non-perturbative  parameters (namely $\mu^2_{\pi}$, $\mu^2_{G}$,  $\rho^3_{D}$ and $\rho^3_{LS}$). Details on the extraction can be find for instance in Ref. \cite{Ricciardi:2019zph}.
The resulting value is 
\beq
|V_{cb}|=(42.19\pm 0.78)\times 10^{-3}
\label{eq:inclvcb_result_ks}
\eeq
\noindent where the quoted uncertainty includes both the experimental and the theoretical uncertainties. It is worth to mention that the latter ones are dominating. In this analysis, the excellent fit quality points toward the validity of the HQE fit, but the small $\chi^2$ per degree of freedoms of $\chi^2/ndf=0.32$, could be a signal of some overestimated theoretical uncertainties, or overestimated correlations between the various moments.

\section{Exclusive $B \to D^{(\ast)}$ decays into light leptons}

One can express the  differential ratios for the semi-leptonic CKM favoured decays $B \to D^{(\ast)} \ell \nu$
 in terms of the  recoil parameter $\omega = p_B \cdot p_{D^{(\ast)}}/m_B \, m_{D^{(\ast)}}$, which corresponds to the energy transferred to the leptonic pair. For negligible lepton masses ($\ell=e, \mu)$, one has
\begin{eqnarray}
&\frac{d\Gamma}{d \omega}(B \rightarrow D^\ast\,\ell {\nu})&
\propto G_F^2    (\omega^2-1)^{\frac{1}{2}}
 |V_{cb}|^2   {\cal F}(\omega)^2
 \nonumber \\
 &\frac{d\Gamma}{d \omega} (B \rightarrow D\,\ell {\nu})&  \propto
G_F^2\,
(\omega^2-1)^{\frac{3}{2}}\,
 |V_{cb}|^2   {\cal G}(\omega)^2
 \label{diffrat}
\end{eqnarray}
In the
heavy quark limit,
both form factors   are related to a single Isgur-Wise
function,  ${\cal F(\omega) }= {\cal G(\omega) } = {\cal  \xi (\omega) }  $, which is
normalized to unity at zero recoil,  that is  ${\cal \xi (\omega}=1) =1 $.
There are
non-perturbative corrections  to this prediction, expressed at the zero-recoil point by the heavy quark symmetry  under the form of
powers of $\Lambda_{QCD}/m$, where $m= m_c$ and $m_b$. Other corrections are perturbatively calculable radiative
corrections from hard gluons and photons.
Latest estimates for zero recoil form factors   come from lattice, and are reported in Table \ref{tab:formf}. We have also listed the form factors for  $ B_s \rightarrow D_s^{(\ast)}  \ell  \nu_\ell $   decays, whose lattice computations are more advantageous, because of the larger mass of the valence $s$ quark compared to $u$ or $d$ quarks. Since the very recent  LHCb measurement \cite{Aaij:2020hsi}, these decays supply a new method for precisely determining $|V_{cb}|$.
\label{subsectionExclusive decays}
\begin{table}[bt!]
\footnotesize
\begin{center}
\vspace*{4mm}
\begin{tabular}{l l c || l c }
\hline 
Collaboration	& Refs. & ${\cal F}(1)$ 	&	 Refs.	&	 ${\cal G}(1) $  \\
\hline
FNAL/MILC  & \cite{Bailey:2014tva} 			& $0.906\pm 0.004 \pm  0.012 $			&	   \cite{Lattice:2015rga} & 
$ 1.054 \pm 0.004 \pm  0.008 $    \\
HPQCD  & \cite{Harrison:2017fmw} &	
$0.895\pm 0.010 \pm  0.024 $			&	  \cite{Na:2015kha}  & $ 1.035\pm 0.040 $       \\
HPQCD  & \cite{McLean:2019sds}	&
$0.914 \pm 0.024 $ &  	&  \\
\hline
 & & ${\cal F}^{ B_s \to D^\ast_s}$(1) &  & ${\cal G}^{ B_s \to D_s}$(1) \\
 \hline
HPQCD & \cite{McLean:2019sds} &  $0.9020 \pm 0.0096 \pm 0.0090$ & \cite{Monahan:2017uby} 
 & $1.068\pm 0.004$
\\
 \hline 
Atoui et al. &  &   & \cite{Atoui:2013zza}
 & $1.052 \pm 0.046$
\\
 \hline 
\end{tabular}
\caption{Latest lattice form factor estimates at zero recoil (From Ref. \cite{Ricciardi:2019zph}).}
\label{tab:formf}
\end{center}
\end{table}

In the computation of the form factors,
the advantage  provided by the heavy quark symmetries  has the hindrance that the differential rates in  \eqref{diffrat} vanish
 at zero recoil.
Thus
 one  needs to extrapolate  the experimental points taken at $\omega \neq 1$ to the zero recoil point $\omega=1$, using a parameterization of the dependence on $\omega$ of the form factors, which introduces additional uncertainties.
Commonly used parameterizations   are the
 CLN
(Caprini-Lellouch-Neubert) \cite{Caprini:1997mu},
 the BGL
(Boyd-Grinstein-Lebed)  \cite{Boyd:1994tt} and the  BCL (Bourrely-Caprini-Lellouch) \cite{Bourrely:2008za} parameterizations.
In all of them,
 $\omega$ is mapped onto a complex variable $z$ via a conformal transformation;
   form factors are written in the form of an
expansion in $z$, which converges rapidly in the kinematical
region of heavy hadron decays. 

Form factor estimates via zero recoil sum rules
  \cite{Faller:2008tr,Gambino:2010bp, Gambino:2012rd} give, in general, relatively higher values of $|V_{cb}|$ and
a theoretical error  more than twice the error in the lattice determinations.  
Recent progress includes
 form factors determinations from $B$ meson light-cone sum rules (LCSR) beyond leading twist in the  $B \to D^{\ast} \ell  \nu$ channel  \cite{Gubernari:2018wyi}.
 
  \section{Exclusive $|V_{cb}|$ determination}
  
\begin{figure}
     \includegraphics[width=1\textwidth]{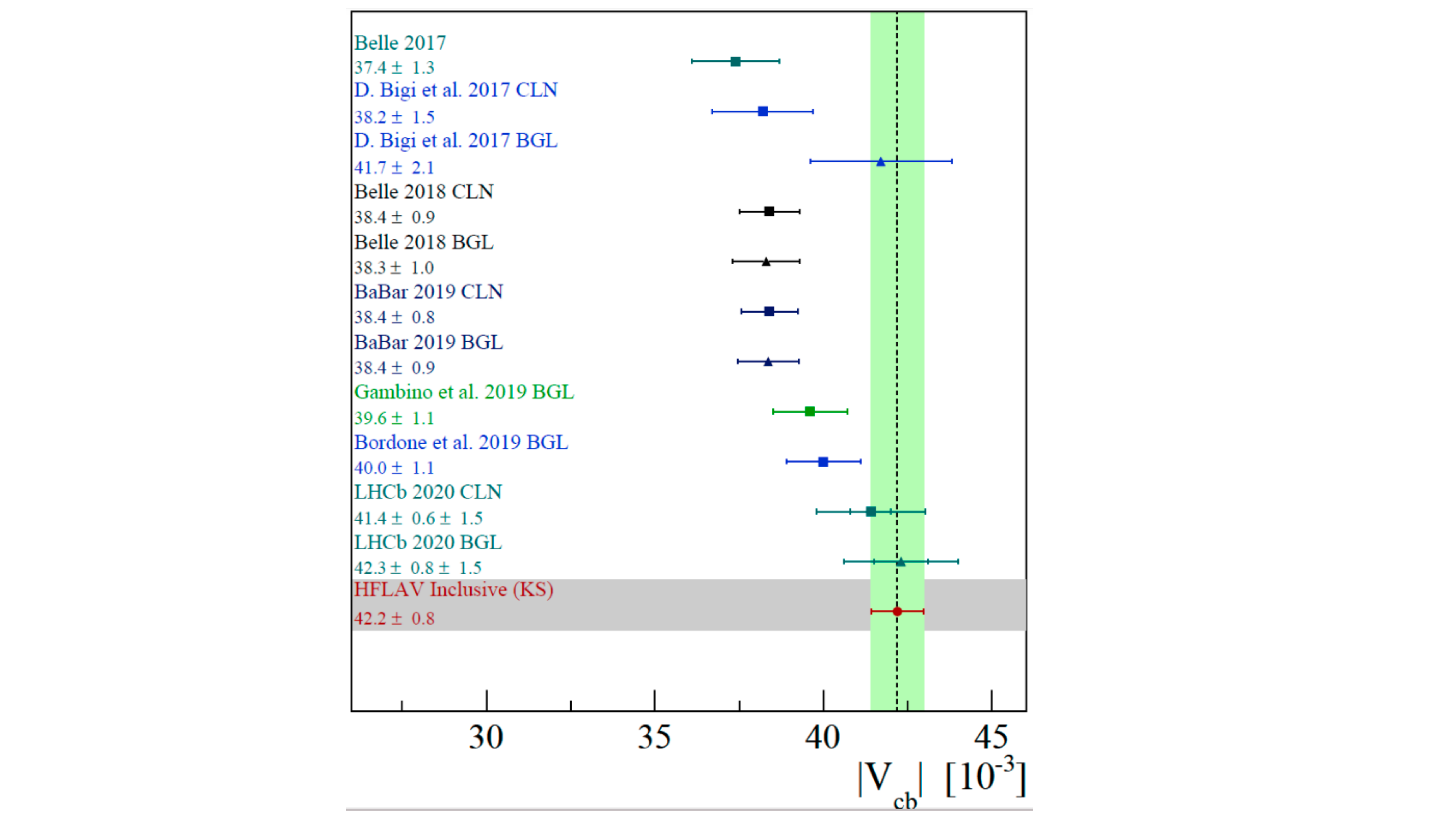}
     \caption{Recent $|V_{cb}|$ determinations (see text).}
     \label{fig1}
     \end{figure}

In 2017,  for the first time, the unfolded
fully-differential decay rate and associated covariance matrix have been published, by the Belle collaboration  \cite{Abdesselam:2017kjf}, prompting  independent determinations of  $|V_{cb}| $ \cite{Grinstein:2017nlq,Bigi:2017njr, Bigi:2017jbd, Jaiswal:2017rve, Bernlochner:2017xyx, Harrison:2017fmw}.
Based on the Belle measurement  \cite{Abdesselam:2017kjf}, the accuracy of the CLN parameterization was questioned in both $B \to D\, l \nu  $ \cite{Bigi:2016mdz} and $B \to D^\ast \, l \nu$ \cite{Bigi:2017njr, Grinstein:2017nlq} channels, in favour of the 
 BGL one \cite{Grinstein:2017nlq,Bigi:2017njr, Bigi:2017jbd}.
Higher central values (closer to the inclusive values)  were found in the latter approach,  and the possibility to have solved the long standing inclusive/exclusive tension was aired.
However, one year later, more data  were provided  by  Belle~\cite{Abdesselam:2018nnh}  and in 2019 by  BaBar~\cite{Dey:2019bgc}.
Both these analyses showed no sign of discrepancy on $|V_{cb}|$ between the BGL and CLN parameterizations, within the uncertainties. Belle also in this case released the data in a format that allows them to be fitted by 
outside groups, prompting a new analysis by some among the authors of  the 2017 fits~\cite{Gambino:2019sif}, which gave consistent results with both CLN and BGL parameterizations,
in different configurations.
        The situation described above is summarized in Fig.~\ref{fig1}~\footnote{For details refer to Ref. \cite{Ricciardi:2019zph}.}.  
        The exclusive values are compared with the inclusive HFLAV average of Eq. \eqref{eq:inclvcb_result_ks}. A recent estimate from  QCD LCSR~\cite{Bordone:2019vic} is reported as well, together with the recent $|V_{cb}|$  determination by LHCb, the first one at a hadron collider and the first one to use $B^0_s$ decays
        \cite{Aaij:2020hsi}.

The ultimate way out of parametrization dependence is to compute the
form factors at nonzero
recoil values. In the case of $B \to D\ell \nu$ decays, the form factors in the unquenched lattice-QCD approximation
have been made available for a range of non-zero recoil momenta since 2015 by the FNAL/MILC  \cite{Lattice:2015rga} and the HPQCD Collaboration~\cite{Na:2015kha}. 
Both estimates  are in good agreement. 
%
Recently, the  HPQCD Collaboration has  presented a lattice QCD determination of the $ B_s \rightarrow D_s  \ell  \nu_\ell $  scalar and vector form factors over the  full  physical  range  of  momentum  transfer \cite{McLean:2019qcx}. 

The LHCb collaboration has measured the ratio of the branching fractions $\Lambda^0_b\to p\mu^-\bar\nu_\mu$ and $\Lambda^0_b\to \Lambda_c^+\mu^-{\bar \nu}_\mu$ \cite{Aaij:2015bfa}, from which they have determined the first direct measurement of the ratio $|V_{ub}|/|V_{cb}|$.
Even if this is not a direct measurement of $|V_{cb}|$, by taking $|V_{ub}|$  from external inputs  it is possible to determine $|V_{cb}|$. 
The LHCb analysis, besides being the first one to use $B$-baryon decays, has paved the way to extract $|V_{ub}|/|V_{cb}|$ from the ratio ${\cal{B}}(B_s^0 \to K^-\mu^+\nu_\mu)/{\cal{B}}(B_s^0\to  D_s^-\mu^+\nu_\mu)$\cite{Aaij:2020nvo}.

\section{Inclusive $|V_{ub}|$ determination}
\label{incl}

In order to extract  $|V_{ub}|$  from semileptonic $B \to X_u \ell \nu$ decays one has to  reduce the $b \to c$ semileptonic background through experimental  cuts. Such cuts enhance the relevance of the so-called threshold region in the phase space, jeopardizing the use of HQE. In order to face this problem, that is absent in the inclusive determination of $|V_{cb}|$,   different theoretical schemes have been devised, which are  tailored
to analyze data in the threshold region,  but  differ
in their treatment of perturbative corrections and the
parametrization of non-perturbative effects.
In  Table \ref{phidectab04} we present the results of four theoretical different approaches, that is ADFR  \cite{Aglietti:2004fz, Aglietti:2006yb,  Aglietti:2007ik}, BLNP
 \cite{Lange:2005yw, Bosch:2004th, Bosch:2004cb}, DGE\cite{Andersen:2005mj} and  GGOU  \cite{Gambino:2007rp},
 as analyzed
by BaBar \cite{Lees:2011fv, Beleno:2013jla}, Belle \cite{Urquijo:2009tp,Cao:2021uwy}  and  HFLAV  \cite{Amhis:2016xyh} collaborations.
  The  most recent (2016) averaged HFLAV determinations \cite{Amhis:2016xyh} are the ones used in the latest (2019) update of PDG \cite{Tanabashi:2018oca}. The results are consistent within the uncertainties.
%
\begin{table}[h]
\centering
\begin{tabular}{lrrrr}
 \hline
 { \color{red}{ Inclusive decays}} &
& {\color{red}{ $ |V_{ub}| \times  10^{3}$}}
  \\
\hline
& { \color{blue}{ ADFR }}    &  { \color{blue}{  BNLP  }}  &   { \color{blue}{  DGE  }}   &  { \color{blue}{   GGOU  }}     \\
\hline
Belle 2020 \cite{Cao:2021uwy} & $4.01^{+ 0.08,+0.15,+0.18}_{-0.08,-0.16,-0.18}$ & $ 4.01^{+ 0.08,+0.15,+0.18}_{-0.08,-0.16,-0.19}$  & $4.12^{+ 0.08,+0.16,+0.11}_{-0.09,-0.16,-0.12}$ &
$4.11^{+ 0.08,+0,16,+0.08}_{-0.09,-0,16,-0.09} $  \\
HFLAV 2016 \cite{Amhis:2016xyh} & $4.08 \pm 0.13^{+ 0.18}_{-0.12}$ & $ 4.44 \pm 0.15^{+0.21}_{-0.22}  $  & $4.52 \pm 0.16^{+ 0.15}_{- 0.16}$ &
$4.52 \pm  0.15^{ + 0.11}_ { -0.14} $  \\
BaBar 2011  \cite{Lees:2011fv} &  $4.29 \pm 0.24^{+0.18}_{-0.19}  $  & $4.28 \pm 0.24^{+0.18}_{-0.20}  $    & $4.40 \pm 0.24^{+0.12}_{-0.13}  $
& $4.35 \pm 0.24^{+0.09}_{-0.10}  $ \\
 Belle 2009 \cite{Urquijo:2009tp} & $4.48 \pm 0.30^{+0.19}_{-0.19}  $ & $ 4.47 \pm 0.27^{+0.19}_{-0.21}  $ &  $4.60 \pm 0.27^{+0.11}_{-0.13}  $ & $4.54 \pm 0.27^{+0.10}_{-0.11}  $ \\
\hline
\end{tabular}
\caption{Status of inclusive $|V_{ub}|$  determinations.}
\label{phidectab04}
\end{table}
The most recent estimates (2020, still preliminary), provided by Belle \cite{Cao:2021uwy}, are also included in Table \ref{phidectab04}:
the  uncertainties are
statistical, systematic and  from the  theory calculation, respectively.
Their arithmetic average of the
results obtained from the  four different QCD predictions gives \cite{Cao:2021uwy}
$
|V_{ub}|=(4.06 \pm 0.09\pm 0.16 \pm 0.15) \times 10^{-3}
$.
This value is smaller than the previous inclusive measurements,  reducing the discrepancy with the exclusive measurement of about 2-3 to  1.4 standard deviations.

\section{Exclusive $|V_{ub}|$ determination}

The CKM-suppressed decay $B \to \pi \ell \nu$ with light final leptons is the typical exclusive channel used to extract $|V_{ub}|$.
It is well-controlled experimentally and several measurements have been performed by both
 BaBar and Belle collaborations \cite{Hokuue:2006nr, Aubert:2006ry, Aubert:2008bf, delAmoSanchez:2010af, Ha:2010rf, Lees:2012vv, Sibidanov:2013rkk}. Since the $u$-quark is not heavy, heavy quark symmetries are not as binding as in 
 $b \to c$ decays.
 Lattice determinations of  the form factors in this channel  have been obtained by  the HPQCD
\cite{Dalgic:2006dt,Colquhoun:2015mfa}, the Fermilab/MILC \cite{Bailey:2008wp, Lattice:2015tia} and the RBC/UKQCD \cite{Flynn:2015mha}  collaborations.
The HFLAV  $|V_{ub}|$  determination comes from a combined fit of a $B \to \pi$ form factor parameterization to theory predictions and the average $q^2$ spectrum in data.
The theory input included in the fit are the results from the FLAG lattice average \cite{Aoki:2016frl} and the LCSR result at $q^2$ = 0 GeV$^2$ \cite{Bharucha:2012wy}.
The results give \cite{Amhis:2019ckw}
$
|V_{ub}| = (3.67 \pm 0.09 \pm 0.12) \times 10^{-3}
$,
where the first error comes from the experiment and the second one from the theory. 

Decay modes with charmless meson states heavier than the charged pion have been studied to a lesser extent; in general they present more challenges in the experimental reconstruction due to possibly higher backgrounds, more complex form-factor dependencies and wider decay widths. 
No lattice unquenched QCD calculation of their form factors is available yet, exception made for preliminary results from the  SPQcdR Collaboration \cite{Abada:2002ie}.
The LCSR computation of form factors
for  $B \to \rho/\omega \, \ell \bar\nu_\ell$ decays has been exploited to estimate $|V_{ub}|$ \cite{Straub:2015ica,Albertus:2014xwa}.  The $B \to \eta^{\prime}$ form factors have also been computed  in the LCSR framework \cite{Ball:2007hb}.
 Measurements of the branching fractions for $B^+ \to \eta/\eta^\prime \ell \nu_\ell$, where $\ell$ stands for either an electron or a muon,  have been reported by  CLEO \cite{Gray:2007pw,Adam:2007pv}, BaBar 
\cite{Aubert:2008bf, Aubert:2008ct, delAmoSanchez:2010zd, Lees:2012vv} and Belle \cite{Beleno:2017cao},



\section{The $|V_{cb}|$ and $|V_{ub}|$ puzzles}

\begin{figure}[t!]
  \centering
  \includegraphics[width=0.9\textwidth]{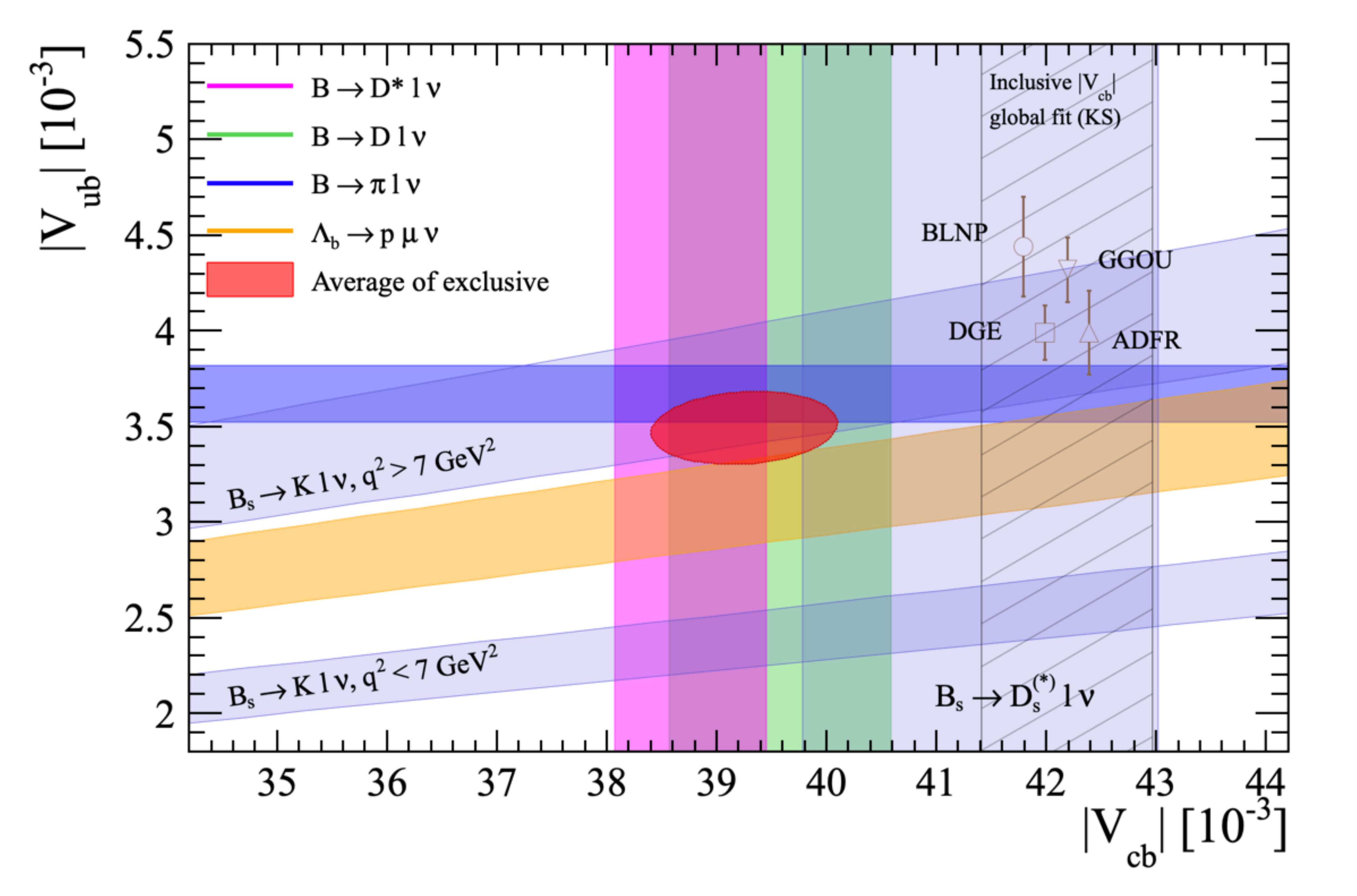}
  \vspace*{-5mm} 
  \caption{$V_{ub}-V_{cb}$ HFLAV averages.}
  \label{fig:vub_vcb_summary}
\end{figure}

The discrepancy between inclusive and exclusive values of $|V_{cb}|$ ($|V_{ub}|$)   is generally referred to as  the $|V_{cb}|$ ($|V_{ub}|$) puzzle. In Fig. \ref{fig:vub_vcb_summary}  HFLAV \cite{Amhis:2019ckw} exclusive and inclusive determinations of $|V_{cb}|$ are summarized and compared with the analogous determinations of $|V_{ub}|$.
The magenta and green vertical bands represent the two different exclusive determinations of $|V_{cb}|$:
 both show a discrepancy with the striped vertical band, which is relative to the $|V_{cb}|$ inclusive determination in the kinetic scheme.  The bands relative to the exclusive $B\to D$ and $B\to D^\ast$ decays are  the HFLAV averages done with the CLN parameterizations. 
Also considering the slightly larger uncertainty associated with the BGL fit, the discrepancy with 
 the inclusive determination remains significant, amounting to about 3$\sigma$. 
 The grey vertical band corresponds to the LHCb result with $B_s\to D_s^{(*)}\mu\nu_\mu$ decays: although affected by large uncertainties, it is
 compatible with both inclusive and exclusive determinations of $|V_{cb}|$.  The oblique bands represent  the constraints on the $|V_{ub}|/|V_{cb}|$ ratio determined  by LHCb  in baryon decays \cite{Aaij:2015bfa} and more recently  in  $B_s^0 \to K^-\mu^+\nu_\mu$ decays \cite{Aaij:2020nvo}.

 In Fig. \ref{fig:vub_vcb_summary}, 
  the blu horizontal band gives the  exclusive  $|V_{ub}|$ determination from semileptonic $B \to \pi$ decay. 
  The values for inclusive $|V_{ub}|$ with different QCD calculations are given by the four points with vertical error bars. The shift along the x-axis of these four points is just arbitrary and has no meaning.  
By considering   their arithmetic average, and the latest Belle results \cite{Cao:2021uwy}, the observed $|V_{ub}|$ discrepancy reduces from  2-3 to 1.4 standard deviations, as discussed in Sect.\ref{incl}.

    It is  also possible to determine $|V_{cb}|$ and $|V_{ub}|$ indirectly, using the CKM unitarity relations together with CP violation and flavour data, excluding direct information on decays, as done by the CKMfitter  \cite{CKMfitter} and by the UTfit Collaborations~\cite{UTfit}. 

Future prospects seems to be  promising.  Belle II is expected to reduce the experimental uncertainty on inclusive and exclusive $|V_{cb}|$ to 1\% and 1.5\%, and on 
inclusive and exclusive $|V_{ub}|$ to 3\% and 2\%, respectively \cite{Kou:2018nap}.
A synergy between theoretical advances, Belle II and the upgraded LHCb may well  say the final word on the present puzzles.

\section*{Acknowledgements}

 The author thanks M. Rotondo for useful discussions. This work received partial financial support  from MIUR under Project No. 2015P5SBHT and from the INFN research initiative ENP.

\bibliographystyle{JHEP2}
\bibliography{main,VxbRef}
\bibliography{main}

\hyphenation{Post-Script Sprin-ger}

\providecommand{\href}[2]{#2}\begingroup\raggedright\begin{thebibliography}{10}

\bibitem{Ricciardi:2019zph}
G.~Ricciardi and M.~Rotondo, \emph{{Determination of the
  Cabibbo-Kobayashi-Maskawa matrix element $|V_{cb}|$}},
  \href{https://doi.org/10.1088/1361-6471/ab9f01}{\emph{J. Phys. G} {\bfseries
  47} (2020) 113001} [\href{https://arxiv.org/abs/1912.09562}{{\ttfamily
  1912.09562}}].

\bibitem{Ricciardi:2017lne}
G.~Ricciardi, \emph{{Semileptonic decays and $|V_{xb}|$ determinations}},
  \href{https://doi.org/10.1051/epjconf/201818202104}{\emph{EPJ Web Conf.}
  {\bfseries 182} (2018) 02104}
  [\href{https://arxiv.org/abs/1712.06988}{{\ttfamily 1712.06988}}].

\bibitem{Koppenburg:2017mad}
S.~Descotes-Genon and P.~Koppenburg, \emph{{The CKM Parameters}},
  \href{https://doi.org/10.1146/annurev-nucl-101916-123109}{\emph{Ann. Rev.
  Nucl. Part. Sci.} {\bfseries 67} (2017) 97}
  [\href{https://arxiv.org/abs/1702.08834}{{\ttfamily 1702.08834}}].

\bibitem{Ricciardi:2016pmh}
G.~Ricciardi, \emph{{Semileptonic and leptonic $B$ decays, circa 2016}},
  \href{https://doi.org/10.1142/S0217732317300051}{\emph{Mod. Phys. Lett.}
  {\bfseries A32} (2017) 1730005}
  [\href{https://arxiv.org/abs/1610.04387}{{\ttfamily 1610.04387}}].

\bibitem{Ricciardi:2014aya}
G.~Ricciardi, \emph{{Status of $|V_{cb}|$ and $|V_{ub}|$ CKM matrix elements}},
  \href{https://doi.org/10.1063/1.4938654}{\emph{AIP Conf. Proc.} {\bfseries
  1701} (2016) 050014} [\href{https://arxiv.org/abs/1412.4288}{{\ttfamily
  1412.4288}}].

\bibitem{Ricciardi:2014iga}
G.~Ricciardi, \emph{{Progress on semi-leptonic $B_{(s)}$ decays}},
  \href{https://doi.org/10.1142/S0217732314300195}{\emph{Mod. Phys. Lett.}
  {\bfseries A29} (2014) 1430019}
  [\href{https://arxiv.org/abs/1403.7750}{{\ttfamily 1403.7750}}].

\bibitem{Melnikov:2008qs}
K.~Melnikov, \emph{{$O(\alpha_2^2)$ corrections to semileptonic decay $ b \to c
  l \bar \nu_l$}},
  \href{https://doi.org/10.1016/j.physletb.2008.07.089}{\emph{Phys. Lett.}
  {\bfseries B666} (2008) 336}
  [\href{https://arxiv.org/abs/0803.0951}{{\ttfamily 0803.0951}}].

\bibitem{Gambino:2011cq}
P.~Gambino, \emph{{B semileptonic moments at NNLO}},
  \href{https://doi.org/10.1007/JHEP09(2011)055}{\emph{JHEP} {\bfseries 09}
  (2011) 055} [\href{https://arxiv.org/abs/1107.3100}{{\ttfamily 1107.3100}}].

\bibitem{Trott:2004xc}
M.~Trott, \emph{{Improving extractions of $|V_{cb}|$ and $m_b$ from the
  hadronic invariant mass moments of semileptonic inclusive B decay}},
  \href{https://doi.org/10.1103/PhysRevD.70.073003}{\emph{Phys. Rev.}
  {\bfseries D70} (2004) 073003}
  [\href{https://arxiv.org/abs/hep-ph/0402120}{{\ttfamily hep-ph/0402120}}].

\bibitem{Aquila:2005hq}
V.~Aquila, P.~Gambino, G.~Ridolfi and N.~Uraltsev, \emph{{Perturbative
  corrections to semileptonic b decay distributions}},
  \href{https://doi.org/10.1016/j.nuclphysb.2005.04.031}{\emph{Nucl. Phys.}
  {\bfseries B719} (2005) 77}
  [\href{https://arxiv.org/abs/hep-ph/0503083}{{\ttfamily hep-ph/0503083}}].

\bibitem{Pak:2008qt}
A.~Pak and A.~Czarnecki, \emph{{Mass effects in muon and semileptonic $ b \to
  c$ decays}},
  \href{https://doi.org/10.1103/PhysRevLett.100.241807}{\emph{Phys. Rev. Lett.}
  {\bfseries 100} (2008) 241807}
  [\href{https://arxiv.org/abs/0803.0960}{{\ttfamily 0803.0960}}].

\bibitem{Pak:2008cp}
A.~Pak and A.~Czarnecki, \emph{{Heavy-to-heavy quark decays at NNLO}},
  \href{https://doi.org/10.1103/PhysRevD.78.114015}{\emph{Phys. Rev.}
  {\bfseries D78} (2008) 114015}
  [\href{https://arxiv.org/abs/0808.3509}{{\ttfamily 0808.3509}}].

\bibitem{Biswas:2009rb}
S.~Biswas and K.~Melnikov, \emph{{Second order QCD corrections to inclusive
  semileptonic $ b \to X_c l \bar \nu$ decays with massless and massive
  lepton}}, \href{https://doi.org/10.1007/JHEP02(2010)089}{\emph{JHEP}
  {\bfseries 02} (2010) 089} [\href{https://arxiv.org/abs/0911.4142}{{\ttfamily
  0911.4142}}].

\bibitem{Benson:2003kp}
D.~Benson, I.I.~Bigi, T.~Mannel and N.~Uraltsev, \emph{{Imprecated, yet
  impeccable: On the theoretical evaluation of $ \Gamma(b \to X_c l \bar
  \nu)$}}, \href{https://doi.org/10.1016/S0550-3213(03)00452-8}{\emph{Nucl.
  Phys.} {\bfseries B665} (2003) 367}
  [\href{https://arxiv.org/abs/hep-ph/0302262}{{\ttfamily hep-ph/0302262}}].

\bibitem{Becher:2007tk}
T.~Becher, H.~Boos and E.~Lunghi, \emph{{Kinetic corrections to $B \to X_{c}
  \ell \bar{\nu}$ at one loop}},
  \href{https://doi.org/10.1088/1126-6708/2007/12/062}{\emph{JHEP} {\bfseries
  12} (2007) 062} [\href{https://arxiv.org/abs/0708.0855}{{\ttfamily
  0708.0855}}].

\bibitem{Alberti:2012dn}
A.~Alberti, T.~Ewerth, P.~Gambino and S.~Nandi, \emph{{Kinetic operator effects
  in $\bar{B}\to X_c l \nu$ at O($\alpha_s$)}},
  \href{https://doi.org/10.1016/j.nuclphysb.2013.01.005}{\emph{Nucl. Phys.}
  {\bfseries B870} (2013) 16}
  [\href{https://arxiv.org/abs/1212.5082}{{\ttfamily 1212.5082}}].

\bibitem{Alberti:2013kxa}
A.~Alberti, P.~Gambino and S.~Nandi, \emph{{Perturbative corrections to power
  suppressed effects in semileptonic B decays}},
  \href{https://doi.org/10.1007/JHEP01(2014)147}{\emph{JHEP} {\bfseries 01}
  (2014) 147} [\href{https://arxiv.org/abs/1311.7381}{{\ttfamily 1311.7381}}].

\bibitem{Mannel:2014xza}
T.~Mannel, A.A.~Pivovarov and D.~Rosenthal, \emph{{Inclusive semileptonic B
  decays from QCD with NLO accuracy for power suppressed terms}},
  \href{https://doi.org/10.1016/j.physletb.2014.12.058}{\emph{Phys. Lett.}
  {\bfseries B741} (2015) 290}
  [\href{https://arxiv.org/abs/1405.5072}{{\ttfamily 1405.5072}}].

\bibitem{Mannel:2015jka}
T.~Mannel, A.A.~Pivovarov and D.~Rosenthal, \emph{{Inclusive weak decays of
  heavy hadrons with power suppressed terms at NLO}},
  \href{https://doi.org/10.1103/PhysRevD.92.054025}{\emph{Phys. Rev.}
  {\bfseries D92} (2015) 054025}
  [\href{https://arxiv.org/abs/1506.08167}{{\ttfamily 1506.08167}}].

\bibitem{Gremm:1996df}
M.~Gremm and A.~Kapustin, \emph{{Order 1/m(b)**3 corrections to B --> X(c)
  lepton anti-neutrino decay and their implication for the measurement of
  Lambda-bar and lambda(1)}},
  \href{https://doi.org/10.1103/PhysRevD.55.6924}{\emph{Phys. Rev.} {\bfseries
  D55} (1997) 6924} [\href{https://arxiv.org/abs/hep-ph/9603448}{{\ttfamily
  hep-ph/9603448}}].

\bibitem{Mannel:2019qel}
T.~Mannel and A.A.~Pivovarov, \emph{{QCD corrections to inclusive heavy hadron
  weak decays at $\Lambda_{\rm QCD}^3 /m_Q^3$}},
  \href{https://arxiv.org/abs/1907.09187}{{\ttfamily 1907.09187}}.

\bibitem{Bigi:2005bh}
I.I.~Bigi, N.~Uraltsev and R.~Zwicky, \emph{{On the nonperturbative charm
  effects in inclusive B ---> X(c) l nu decays}},
  \href{https://doi.org/10.1140/epjc/s10052-007-0216-8}{\emph{Eur. Phys. J.}
  {\bfseries C50} (2007) 539}
  [\href{https://arxiv.org/abs/hep-ph/0511158}{{\ttfamily hep-ph/0511158}}].

\bibitem{Breidenbach:2008ua}
C.~Breidenbach, T.~Feldmann, T.~Mannel and S.~Turczyk, \emph{{On the Role of
  'Intrinsic Charm' in Semi-Leptonic B-Meson Decays}},
  \href{https://doi.org/10.1103/PhysRevD.78.014022}{\emph{Phys. Rev.}
  {\bfseries D78} (2008) 014022}
  [\href{https://arxiv.org/abs/0805.0971}{{\ttfamily 0805.0971}}].

\bibitem{Bigi:2009ym}
I.~Bigi, T.~Mannel, S.~Turczyk and N.~Uraltsev, \emph{{The Two Roads to
  'Intrinsic Charm' in B Decays}},
  \href{https://doi.org/10.1007/JHEP04(2010)073}{\emph{JHEP} {\bfseries 04}
  (2010) 073} [\href{https://arxiv.org/abs/0911.3322}{{\ttfamily 0911.3322}}].

\bibitem{Dassinger:2006md}
B.M.~Dassinger, T.~Mannel and S.~Turczyk, \emph{{Inclusive semi-leptonic B
  decays to order 1 / m(b)**4}},
  \href{https://doi.org/10.1088/1126-6708/2007/03/087}{\emph{JHEP} {\bfseries
  03} (2007) 087} [\href{https://arxiv.org/abs/hep-ph/0611168}{{\ttfamily
  hep-ph/0611168}}].

\bibitem{Mannel:2010wj}
T.~Mannel, S.~Turczyk and N.~Uraltsev, \emph{{Higher Order Power Corrections in
  Inclusive B Decays}},
  \href{https://doi.org/10.1007/JHEP11(2010)109}{\emph{JHEP} {\bfseries 11}
  (2010) 109} [\href{https://arxiv.org/abs/1009.4622}{{\ttfamily 1009.4622}}].

\bibitem{Heinonen:2014dxa}
J.~Heinonen and T.~Mannel, \emph{{Improved Estimates for the Parameters of the
  Heavy Quark Expansion}},
  \href{https://doi.org/10.1016/j.nuclphysb.2014.09.017}{\emph{Nucl. Phys.}
  {\bfseries B889} (2014) 46}
  [\href{https://arxiv.org/abs/1407.4384}{{\ttfamily 1407.4384}}].

\bibitem{Amhis:2019ckw}
{\scshape HFLAV} collaboration, \emph{{Averages of $b$-hadron, $c$-hadron, and
  $\tau$-lepton properties as of 2018 (updates on website)}},
  \href{https://arxiv.org/abs/1909.12524}{{\ttfamily 1909.12524}}.

\bibitem{Aaij:2020hsi}
{\scshape LHCb} collaboration, \emph{{Measurement of $|V_{cb}|$ with $B_s^0 \to
  D_s^{(*)-} \mu^+ \nu_{\mu}$ decays}},
  \href{https://doi.org/10.1103/PhysRevD.101.072004}{\emph{Phys. Rev. D}
  {\bfseries 101} (2020) 072004}
  [\href{https://arxiv.org/abs/2001.03225}{{\ttfamily 2001.03225}}].

\bibitem{Bailey:2014tva}
{\scshape Fermilab Lattice, MILC} collaboration, \emph{{Update of $|V_{cb}|$
  from the $\bar{B}\to D^*\ell\bar{\nu}$ form factor at zero recoil with
  three-flavor lattice QCD}},
  \href{https://doi.org/10.1103/PhysRevD.89.114504}{\emph{Phys. Rev.}
  {\bfseries D89} (2014) 114504}
  [\href{https://arxiv.org/abs/1403.0635}{{\ttfamily 1403.0635}}].

\bibitem{Lattice:2015rga}
{\scshape MILC} collaboration, \emph{{$B\to D\ell\nu$ form factors at nonzero
  recoil and |V$_{cb}$| from 2+1-flavor lattice QCD}},
  \href{https://doi.org/10.1103/PhysRevD.92.034506}{\emph{Phys. Rev.}
  {\bfseries D92} (2015) 034506}
  [\href{https://arxiv.org/abs/1503.07237}{{\ttfamily 1503.07237}}].

\bibitem{Harrison:2017fmw}
{\scshape HPQCD} collaboration, \emph{{Lattice QCD calculation of the
  ${{B}_{(s)}\to D_{(s)}^{*}\ell{\nu}}$ form factors at zero recoil and
  implications for ${|V_{cb}|}$}},
  \href{https://doi.org/10.1103/PhysRevD.97.054502}{\emph{Phys. Rev.}
  {\bfseries D97} (2018) 054502}
  [\href{https://arxiv.org/abs/1711.11013}{{\ttfamily 1711.11013}}].

\bibitem{Na:2015kha}
{\scshape HPQCD} collaboration, \emph{{$B \rightarrow D l \nu$ form factors at
  nonzero recoil and extraction of $|V_{cb}|$}},
  \href{https://doi.org/10.1103/PhysRevD.93.119906,
  10.1103/PhysRevD.92.054510}{\emph{Phys. Rev.} {\bfseries D92} (2015) 054510}
  [\href{https://arxiv.org/abs/1505.03925}{{\ttfamily 1505.03925}}].

\bibitem{McLean:2019sds}
E.~McLean, C.T.H.~Davies, A.T.~Lytle and J.~Koponen, \emph{{Lattice QCD form
  factor for $B_s\to D_s^* l\nu$ at zero recoil with non-perturbative current
  renormalisation}},
  \href{https://doi.org/10.1103/PhysRevD.99.114512}{\emph{Phys. Rev.}
  {\bfseries D99} (2019) 114512}
  [\href{https://arxiv.org/abs/1904.02046}{{\ttfamily 1904.02046}}].

\bibitem{Monahan:2017uby}
C.J.~Monahan, H.~Na, C.M.~Bouchard, G.P.~Lepage and J.~Shigemitsu, \emph{{$B_s
  \to D_s \ell \nu$ Form Factors and the Fragmentation Fraction Ratio
  $f_s/f_d$}}, \href{https://doi.org/10.1103/PhysRevD.95.114506}{\emph{Phys.
  Rev.} {\bfseries D95} (2017) 114506}
  [\href{https://arxiv.org/abs/1703.09728}{{\ttfamily 1703.09728}}].

\bibitem{Atoui:2013zza}
M.~Atoui, V.~Morénas, D.~Bečirevic and F.~Sanfilippo, \emph{{$B_{s} \to
  D_{s}\ell\nu_\ell$ near zero recoil in and beyond the Standard Model}},
  \href{https://doi.org/10.1140/epjc/s10052-014-2861-z}{\emph{Eur. Phys. J.}
  {\bfseries C74} (2014) 2861}
  [\href{https://arxiv.org/abs/1310.5238}{{\ttfamily 1310.5238}}].

\bibitem{Caprini:1997mu}
I.~Caprini, L.~Lellouch and M.~Neubert, \emph{{Dispersive bounds on the shape
  of anti-B ---> D(*) lepton anti-neutrino form-factors}},
  \href{https://doi.org/10.1016/S0550-3213(98)00350-2}{\emph{Nucl. Phys.}
  {\bfseries B530} (1998) 153}
  [\href{https://arxiv.org/abs/hep-ph/9712417}{{\ttfamily hep-ph/9712417}}].

\bibitem{Boyd:1994tt}
C.G.~Boyd, B.~Grinstein and R.F.~Lebed, \emph{{Constraints on form-factors for
  exclusive semileptonic heavy to light meson decays}},
  \href{https://doi.org/10.1103/PhysRevLett.74.4603}{\emph{Phys. Rev. Lett.}
  {\bfseries 74} (1995) 4603}
  [\href{https://arxiv.org/abs/hep-ph/9412324}{{\ttfamily hep-ph/9412324}}].

\bibitem{Bourrely:2008za}
C.~Bourrely, I.~Caprini and L.~Lellouch, \emph{{Model-independent description
  of B ---> pi l nu decays and a determination of |V(ub)|}},
  \href{https://doi.org/10.1103/PhysRevD.82.099902,
  10.1103/PhysRevD.79.013008}{\emph{Phys. Rev.} {\bfseries D79} (2009) 013008}
  [\href{https://arxiv.org/abs/0807.2722}{{\ttfamily 0807.2722}}].

\bibitem{Faller:2008tr}
S.~Faller, A.~Khodjamirian, C.~Klein and T.~Mannel, \emph{{B ---> D(*) Form
  Factors from QCD Light-Cone Sum Rules}},
  \href{https://doi.org/10.1140/epjc/s10052-009-0968-4}{\emph{Eur. Phys. J.}
  {\bfseries C60} (2009) 603}
  [\href{https://arxiv.org/abs/0809.0222}{{\ttfamily 0809.0222}}].

\bibitem{Gambino:2010bp}
P.~Gambino, T.~Mannel and N.~Uraltsev, \emph{{B -> D* at zero recoil
  revisited}}, \href{https://doi.org/10.1103/PhysRevD.81.113002}{\emph{Phys.
  Rev.} {\bfseries D81} (2010) 113002}
  [\href{https://arxiv.org/abs/1004.2859}{{\ttfamily 1004.2859}}].

\bibitem{Gambino:2012rd}
P.~Gambino, T.~Mannel and N.~Uraltsev, \emph{{B-> D* Zero-Recoil Formfactor and
  the Heavy Quark Expansion in QCD: A Systematic Study}},
  \href{https://doi.org/10.1007/JHEP10(2012)169}{\emph{JHEP} {\bfseries 10}
  (2012) 169} [\href{https://arxiv.org/abs/1206.2296}{{\ttfamily 1206.2296}}].

\bibitem{Gubernari:2018wyi}
N.~Gubernari, A.~Kokulu and D.~van Dyk, \emph{{$B\to P$ and $B\to V$ Form
  Factors from $B$-Meson Light-Cone Sum Rules beyond Leading Twist}},
  \href{https://doi.org/10.1007/JHEP01(2019)150}{\emph{JHEP} {\bfseries 01}
  (2019) 150} [\href{https://arxiv.org/abs/1811.00983}{{\ttfamily
  1811.00983}}].

\bibitem{Abdesselam:2017kjf}
{\scshape Belle} collaboration, \emph{{Precise determination of the CKM matrix
  element $\left| V_{cb}\right|$ with $\bar B^0 \to D^{*\,+} \, \ell^- \, \bar
  \nu_\ell$ decays with hadronic tagging at Belle}},
  \href{https://arxiv.org/abs/1702.01521}{{\ttfamily 1702.01521}}.

\bibitem{Grinstein:2017nlq}
B.~Grinstein and A.~Kobach, \emph{{Model-Independent Extraction of $|V_{cb}|$
  from $\bar{B}\rightarrow D^* \ell \overline{\nu}$}},
  \href{https://doi.org/10.1016/j.physletb.2017.05.078}{\emph{Phys. Lett.}
  {\bfseries B771} (2017) 359}
  [\href{https://arxiv.org/abs/1703.08170}{{\ttfamily 1703.08170}}].

\bibitem{Bigi:2017njr}
D.~Bigi, P.~Gambino and S.~Schacht, \emph{{A fresh look at the determination of
  $|V_{cb}|$ from $B\to D^{*} \ell \nu$}},
  \href{https://doi.org/10.1016/j.physletb.2017.04.022}{\emph{Phys. Lett.}
  {\bfseries B769} (2017) 441}
  [\href{https://arxiv.org/abs/1703.06124}{{\ttfamily 1703.06124}}].

\bibitem{Bigi:2017jbd}
D.~Bigi, P.~Gambino and S.~Schacht, \emph{{$R(D^*)$, $|V_{cb}|$, and the Heavy
  Quark Symmetry relations between form factors}},
  \href{https://doi.org/10.1007/JHEP11(2017)061}{\emph{JHEP} {\bfseries 11}
  (2017) 061} [\href{https://arxiv.org/abs/1707.09509}{{\ttfamily
  1707.09509}}].

\bibitem{Jaiswal:2017rve}
S.~Jaiswal, S.~Nandi and S.K.~Patra, \emph{{Extraction of $|V_{cb}|$ from $B\to
  D^{(*)}\ell\nu_\ell$ and the Standard Model predictions of $R(D^{(*)})$}},
  \href{https://doi.org/10.1007/JHEP12(2017)060}{\emph{JHEP} {\bfseries 12}
  (2017) 060} [\href{https://arxiv.org/abs/1707.09977}{{\ttfamily
  1707.09977}}].

\bibitem{Bernlochner:2017xyx}
F.U.~Bernlochner, Z.~Ligeti, M.~Papucci and D.J.~Robinson, \emph{{Tensions and
  correlations in $|V_{cb}|$ determinations}},
  \href{https://doi.org/10.1103/PhysRevD.96.091503}{\emph{Phys. Rev.}
  {\bfseries D96} (2017) 091503}
  [\href{https://arxiv.org/abs/1708.07134}{{\ttfamily 1708.07134}}].

\bibitem{Bigi:2016mdz}
D.~Bigi and P.~Gambino, \emph{{Revisiting $B\to D \ell \nu$}},
  \href{https://doi.org/10.1103/PhysRevD.94.094008}{\emph{Phys. Rev.}
  {\bfseries D94} (2016) 094008}
  [\href{https://arxiv.org/abs/1606.08030}{{\ttfamily 1606.08030}}].

\bibitem{Abdesselam:2018nnh}
{\scshape Belle} collaboration, \emph{{Measurement of CKM Matrix Element
  $|V_{cb}|$ from $\bar{B} \to D^{*+} \ell^{-} \bar{\nu}_\ell$}},
  \href{https://arxiv.org/abs/1809.03290}{{\ttfamily 1809.03290}}.

\bibitem{Dey:2019bgc}
{\scshape BaBar} collaboration, \emph{{Extraction of form Factors from a
  Four-Dimensional Angular Analysis of $\overline{B} \rightarrow D^\ast \ell^-
  \overline{\nu}_\ell$}},
  \href{https://doi.org/10.1103/PhysRevLett.123.091801}{\emph{Phys. Rev. Lett.}
  {\bfseries 123} (2019) 091801}
  [\href{https://arxiv.org/abs/1903.10002}{{\ttfamily 1903.10002}}].

\bibitem{Gambino:2019sif}
P.~Gambino, M.~Jung and S.~Schacht, \emph{{The $V_{cb}$ puzzle: An update}},
  \href{https://doi.org/10.1016/j.physletb.2019.06.039}{\emph{Phys. Lett.}
  {\bfseries B795} (2019) 386}
  [\href{https://arxiv.org/abs/1905.08209}{{\ttfamily 1905.08209}}].

\bibitem{Bordone:2019vic}
M.~Bordone, M.~Jung and D.~van Dyk, \emph{{Theory determination of $\bar{B}\to
  D^{(*)}\ell^-\bar\nu$ form factors at $\mathcal{O}(1/m_c^2)$}},
  \href{https://doi.org/10.1140/epjc/s10052-020-7616-4}{\emph{Eur. Phys. J. C}
  {\bfseries 80} (2020) 74} [\href{https://arxiv.org/abs/1908.09398}{{\ttfamily
  1908.09398}}].

\bibitem{McLean:2019qcx}
E.~McLean, C.T.H.~Davies, J.~Koponen and A.T.~Lytle, \emph{{$B_s\to D_s
  \ell\nu$ Form Factors for the full $q^2$ range from Lattice QCD with
  non-perturbatively normalized currents}},
  \href{https://arxiv.org/abs/1906.00701}{{\ttfamily 1906.00701}}.

\bibitem{Aaij:2015bfa}
{\scshape LHCb} collaboration, \emph{{Determination of the quark coupling
  strength $|V_{ub}|$ using baryonic decays}},
  \href{https://doi.org/10.1038/nphys3415}{\emph{Nature Phys.} {\bfseries 11}
  (2015) 743} [\href{https://arxiv.org/abs/1504.01568}{{\ttfamily
  1504.01568}}].

\bibitem{Aaij:2020nvo}
{\scshape LHCb} collaboration, \emph{{First observation of the decay $B_s^0 \to
  K^-\mu^+\nu_\mu$ and measurement of $|V_{ub}|/|V_{cb}|$}},
  \href{https://arxiv.org/abs/2012.05143}{{\ttfamily 2012.05143}}.

\bibitem{Aglietti:2004fz}
U.~Aglietti and G.~Ricciardi, \emph{{A Model for next-to-leading order
  threshold resummed form-factors}},
  \href{https://doi.org/10.1103/PhysRevD.70.114008}{\emph{Phys. Rev.}
  {\bfseries D70} (2004) 114008}
  [\href{https://arxiv.org/abs/hep-ph/0407225}{{\ttfamily hep-ph/0407225}}].

\bibitem{Aglietti:2006yb}
U.~Aglietti, G.~Ferrera and G.~Ricciardi, \emph{{Semi-Inclusive B Decays and a
  Model for Soft-Gluon Effects}},
  \href{https://doi.org/10.1016/j.nuclphysb.2007.01.014}{\emph{Nucl. Phys.}
  {\bfseries B768} (2007) 85}
  [\href{https://arxiv.org/abs/hep-ph/0608047}{{\ttfamily hep-ph/0608047}}].

\bibitem{Aglietti:2007ik}
U.~Aglietti, F.~Di~Lodovico, G.~Ferrera and G.~Ricciardi, \emph{{Inclusive
  measure of |V(ub)| with the analytic coupling model}},
  \href{https://doi.org/10.1140/epjc/s10052-008-0817-x}{\emph{Eur. Phys. J.}
  {\bfseries C59} (2009) 831}
  [\href{https://arxiv.org/abs/0711.0860}{{\ttfamily 0711.0860}}].

\bibitem{Lange:2005yw}
B.O.~Lange, M.~Neubert and G.~Paz, \emph{{Theory of charmless inclusive B
  decays and the extraction of V(ub)}},
  \href{https://doi.org/10.1103/PhysRevD.72.073006}{\emph{Phys. Rev.}
  {\bfseries D72} (2005) 073006}
  [\href{https://arxiv.org/abs/hep-ph/0504071}{{\ttfamily hep-ph/0504071}}].

\bibitem{Bosch:2004th}
S.W.~Bosch, B.O.~Lange, M.~Neubert and G.~Paz, \emph{{Factorization and shape
  function effects in inclusive B meson decays}},
  \href{https://doi.org/10.1016/j.nuclphysb.2004.07.041}{\emph{Nucl. Phys.}
  {\bfseries B699} (2004) 335}
  [\href{https://arxiv.org/abs/hep-ph/0402094}{{\ttfamily hep-ph/0402094}}].

\bibitem{Bosch:2004cb}
S.W.~Bosch, M.~Neubert and G.~Paz, \emph{{Subleading shape functions in
  inclusive B decays}},
  \href{https://doi.org/10.1088/1126-6708/2004/11/073}{\emph{JHEP} {\bfseries
  11} (2004) 073} [\href{https://arxiv.org/abs/hep-ph/0409115}{{\ttfamily
  hep-ph/0409115}}].

\bibitem{Andersen:2005mj}
J.R.~Andersen and E.~Gardi, \emph{{Inclusive spectra in charmless semileptonic
  B decays by dressed gluon exponentiation}},
  \href{https://doi.org/10.1088/1126-6708/2006/01/097}{\emph{JHEP} {\bfseries
  01} (2006) 097} [\href{https://arxiv.org/abs/hep-ph/0509360}{{\ttfamily
  hep-ph/0509360}}].

\bibitem{Gambino:2007rp}
P.~Gambino, P.~Giordano, G.~Ossola and N.~Uraltsev, \emph{{Inclusive
  semileptonic B decays and the determination of |V(ub)|}},
  \href{https://doi.org/10.1088/1126-6708/2007/10/058}{\emph{JHEP} {\bfseries
  10} (2007) 058} [\href{https://arxiv.org/abs/0707.2493}{{\ttfamily
  0707.2493}}].

\bibitem{Lees:2011fv}
{\scshape BaBar} collaboration, \emph{{Study of $\bar{B}\to X_u \ell \bar{\nu}$
  decays in $B\bar{B}$ events tagged by a fully reconstructed B-meson decay and
  determination of $|V_{ub}|$}},
  \href{https://doi.org/10.1103/PhysRevD.86.032004}{\emph{Phys.Rev.} {\bfseries
  D86} (2012) 032004} [\href{https://arxiv.org/abs/1112.0702}{{\ttfamily
  1112.0702}}].

\bibitem{Beleno:2013jla}
C.~Beleno, \emph{{Charmless Semileptonic B Decays at e+e- Colliders}},  in
  \emph{{Proceedings, 11th Conference on Flavor Physics and CP Violation (FPCP
  2013), May 20-24, Armacao dos Buzios, Rio de Janeiro, Brazil}}, 2013
  [\href{https://arxiv.org/abs/1307.8285}{{\ttfamily 1307.8285}}].

\bibitem{Urquijo:2009tp}
{\scshape Belle} collaboration, \emph{{Measurement of $|V_{ub}|$ From Inclusive
  Charmless Semileptonic B Decays}},
  \href{https://doi.org/10.1103/PhysRevLett.104.021801}{\emph{Phys.Rev.Lett.}
  {\bfseries 104} (2010) 021801}
  [\href{https://arxiv.org/abs/0907.0379}{{\ttfamily 0907.0379}}].

\bibitem{Cao:2021uwy}
{\scshape Belle} collaboration, \emph{{New results on inclusive $B\to X_{u}
  \ell \nu$ decay from the Belle experiment}},  in \emph{{40th International
  Conference on High Energy Physics}}, 1, 2021
  [\href{https://arxiv.org/abs/2101.04512}{{\ttfamily 2101.04512}}].

\bibitem{Amhis:2016xyh}
{\scshape HFLAV} collaboration, \emph{{Averages of $b$-hadron, $c$-hadron, and
  $\tau$-lepton properties as of summer 2016}},
  \href{https://doi.org/10.1140/epjc/s10052-017-5058-4}{\emph{Eur. Phys. J.}
  {\bfseries C77} (2017) 895}
  [\href{https://arxiv.org/abs/1612.07233}{{\ttfamily 1612.07233}}].

\bibitem{Tanabashi:2018oca}
{\scshape Particle Data Group} collaboration, \emph{{Review of Particle
  Physics}}, \href{https://doi.org/10.1103/PhysRevD.98.030001}{\emph{Phys.
  Rev.} {\bfseries D98} (2018) 030001}.

\bibitem{Hokuue:2006nr}
{\scshape Belle} collaboration, \emph{{Measurements of branching fractions and
  q**2 distributions for B ---> pi l nu and B ---> rho l nu decays with B --->
  D(*) l nu decay tagging}},
  \href{https://doi.org/10.1016/j.physletb.2007.02.067}{\emph{Phys. Lett.}
  {\bfseries B648} (2007) 139}
  [\href{https://arxiv.org/abs/hep-ex/0604024}{{\ttfamily hep-ex/0604024}}].

\bibitem{Aubert:2006ry}
{\scshape BaBar} collaboration, \emph{{Measurement of the $B \to \pi \ell \nu$
  Branching Fraction and Determination of $|V_{ub}|$ with Tagged $B$ Mesons}},
  \href{https://doi.org/10.1103/PhysRevLett.97.211801}{\emph{Phys. Rev. Lett.}
  {\bfseries 97} (2006) 211801}
  [\href{https://arxiv.org/abs/hep-ex/0607089}{{\ttfamily hep-ex/0607089}}].

\bibitem{Aubert:2008bf}
{\scshape BaBar} collaboration, \emph{{Measurements of $B \to \{\pi, \eta,
  \eta^\prime\} \ell \nu_{\ell}$ Branching Fractions and Determination of
  $|V_{ub}|$ with Semileptonically Tagged $B$ Mesons}},
  \href{https://doi.org/10.1103/PhysRevLett.101.081801}{\emph{Phys. Rev. Lett.}
  {\bfseries 101} (2008) 081801}
  [\href{https://arxiv.org/abs/0805.2408}{{\ttfamily 0805.2408}}].

\bibitem{delAmoSanchez:2010af}
{\scshape BaBar} collaboration, \emph{{Study of $B \to \pi \ell \nu$ and $B \to
  \rho \ell \nu$ Decays and Determination of $|V_{ub}|$}},
  \href{https://doi.org/10.1103/PhysRevD.83.032007}{\emph{Phys.Rev.} {\bfseries
  D83} (2011) 032007} [\href{https://arxiv.org/abs/1005.3288}{{\ttfamily
  1005.3288}}].

\bibitem{Ha:2010rf}
{\scshape Belle} collaboration, \emph{{Measurement of the decay
  $B^0\to\pi^-\ell^+\nu$ and determination of $|V_{ub}|$}},
  \href{https://doi.org/10.1103/PhysRevD.83.071101}{\emph{Phys.Rev.} {\bfseries
  D83} (2011) 071101} [\href{https://arxiv.org/abs/1012.0090}{{\ttfamily
  1012.0090}}].

\bibitem{Lees:2012vv}
{\scshape BaBar} collaboration, \emph{{Branching fraction and form-factor shape
  measurements of exclusive charmless semileptonic B decays, and determination
  of $|V_{ub}|$}},
  \href{https://doi.org/10.1103/PhysRevD.86.092004}{\emph{Phys.Rev.} {\bfseries
  D86} (2012) 092004} [\href{https://arxiv.org/abs/1208.1253}{{\ttfamily
  1208.1253}}].

\bibitem{Sibidanov:2013rkk}
{\scshape Belle} collaboration, \emph{{Study of Exclusive $B \to X_u \ell \nu$
  Decays and Extraction of $\|V_{ub}\|$ using Full Reconstruction Tagging at
  the Belle Experiment}},
  \href{https://doi.org/10.1103/PhysRevD.88.032005}{\emph{Phys.Rev.} {\bfseries
  D88} (2013) 032005} [\href{https://arxiv.org/abs/1306.2781}{{\ttfamily
  1306.2781}}].

\bibitem{Dalgic:2006dt}
E.~Dalgic, A.~Gray, M.~Wingate, C.T.H.~Davies, G.P.~Lepage and J.~Shigemitsu,
  \emph{{B meson semileptonic form-factors from unquenched lattice QCD}},
  \href{https://doi.org/10.1103/PhysRevD.75.119906,
  10.1103/PhysRevD.73.074502}{\emph{Phys. Rev.} {\bfseries D73} (2006) 074502}
  [\href{https://arxiv.org/abs/hep-lat/0601021}{{\ttfamily hep-lat/0601021}}].

\bibitem{Colquhoun:2015mfa}
B.~Colquhoun, R.J.~Dowdall, J.~Koponen, C.T.H.~Davies and G.P.~Lepage, \emph{{$
  B \to \pi l \nu$ at zero recoil from lattice QCD with physical u/d quarks}},
  \href{https://doi.org/10.1103/PhysRevD.93.034502}{\emph{Phys. Rev.}
  {\bfseries D93} (2016) 034502}
  [\href{https://arxiv.org/abs/1510.07446}{{\ttfamily 1510.07446}}].

\bibitem{Bailey:2008wp}
J.A.~Bailey, C.~Bernard, C.E.~DeTar, M.~Di~Pierro, A.~El-Khadra et~al.,
  \emph{{The $B \to \pi \ell \nu$ semileptonic form factor from three-flavor
  lattice QCD: A Model-independent determination of $|V_{ub}|$}},
  \href{https://doi.org/10.1103/PhysRevD.79.054507}{\emph{Phys.Rev.} {\bfseries
  D79} (2009) 054507} [\href{https://arxiv.org/abs/0811.3640}{{\ttfamily
  0811.3640}}].

\bibitem{Lattice:2015tia}
{\scshape Fermilab Lattice, MILC} collaboration, \emph{{$|V_{ub}|$ from
  $B\to\pi\ell\nu$ decays and (2+1)-flavor lattice QCD}},
  \href{https://doi.org/10.1103/PhysRevD.92.014024}{\emph{Phys. Rev.}
  {\bfseries D92} (2015) 014024}
  [\href{https://arxiv.org/abs/1503.07839}{{\ttfamily 1503.07839}}].

\bibitem{Flynn:2015mha}
J.M.~Flynn, T.~Izubuchi, T.~Kawanai, C.~Lehner, A.~Soni, R.S.~Van~de Water
  et~al., \emph{{$B \to \pi \ell \nu$ and $B_s \to K \ell \nu$ form factors and
  $|V_{ub}|$ from 2+1-flavor lattice QCD with domain-wall light quarks and
  relativistic heavy quarks}},
  \href{https://doi.org/10.1103/PhysRevD.91.074510}{\emph{Phys. Rev.}
  {\bfseries D91} (2015) 074510}
  [\href{https://arxiv.org/abs/1501.05373}{{\ttfamily 1501.05373}}].

\bibitem{Aoki:2016frl}
S.~Aoki et~al., \emph{{Review of lattice results concerning low-energy particle
  physics}}, \href{https://doi.org/10.1140/epjc/s10052-016-4509-7}{\emph{Eur.
  Phys. J.} {\bfseries C77} (2017) 112}
  [\href{https://arxiv.org/abs/1607.00299}{{\ttfamily 1607.00299}}].

\bibitem{Bharucha:2012wy}
A.~Bharucha, \emph{{Two-loop Corrections to the B to pi Form Factor from QCD
  Sum Rules on the Light-Cone and |V(ub)|}},
  \href{https://doi.org/10.1007/JHEP05(2012)092}{\emph{JHEP} {\bfseries 1205}
  (2012) 092} [\href{https://arxiv.org/abs/1203.1359}{{\ttfamily 1203.1359}}].

\bibitem{Abada:2002ie}
{\scshape SPQcdR} collaboration, \emph{{Heavy to light vector meson
  semileptonic decays}},
  \href{https://doi.org/10.1016/S0920-5632(03)01643-8}{\emph{Nucl. Phys. B
  Proc. Suppl.} {\bfseries 119} (2003) 625}
  [\href{https://arxiv.org/abs/hep-lat/0209116}{{\ttfamily hep-lat/0209116}}].

\bibitem{Straub:2015ica}
A.~Bharucha, D.M.~Straub and R.~Zwicky, \emph{{$B\to V l^+ l^-$ in the Standard
  Model from light-cone sum rules}},
  \href{https://doi.org/10.1007/JHEP08(2016)098}{\emph{JHEP} {\bfseries 08}
  (2016) 098} [\href{https://arxiv.org/abs/1503.05534}{{\ttfamily
  1503.05534}}].

\bibitem{Albertus:2014xwa}
C.~Albertus, E.~Hern\'andez and J.~Nieves,
  \emph{{B\textrightarrow{}\ensuremath{\rho} semileptonic decays and
  |V$_ub$|}}, \href{https://doi.org/10.1103/PhysRevD.90.013017}{\emph{Phys.
  Rev. D} {\bfseries 90} (2014) 013017}
  [\href{https://arxiv.org/abs/1406.7782}{{\ttfamily 1406.7782}}].

\bibitem{Gray:2007pw}
{\scshape CLEO} collaboration, \emph{{A Study of Exclusive Charmless
  Semileptonic B Decays and Extraction of $|V_{ub}|$ at CLEO}},
  \href{https://doi.org/10.1103/PhysRevD.76.012007}{\emph{Phys. Rev. D}
  {\bfseries 76} (2007) 012007}
  [\href{https://arxiv.org/abs/hep-ex/0703042}{{\ttfamily hep-ex/0703042}}].

\bibitem{Adam:2007pv}
{\scshape CLEO} collaboration, \emph{{A Study of Exclusive Charmless
  Semileptonic B Decay and |V(ub)|}},
  \href{https://doi.org/10.1103/PhysRevLett.99.041802}{\emph{Phys. Rev. Lett.}
  {\bfseries 99} (2007) 041802}
  [\href{https://arxiv.org/abs/hep-ex/0703041}{{\ttfamily hep-ex/0703041}}].

\bibitem{Aubert:2008ct}
{\scshape BaBar} collaboration, \emph{{Measurement of the $B^{+} \to \omega
  \ell^{+} \nu$ and $B^{+}$ -- $\to \eta \ell^{+} \nu$ Branching Fractions}},
  \href{https://doi.org/10.1103/PhysRevD.79.052011}{\emph{Phys. Rev. D}
  {\bfseries 79} (2009) 052011}
  [\href{https://arxiv.org/abs/0808.3524}{{\ttfamily 0808.3524}}].

\bibitem{delAmoSanchez:2010zd}
{\scshape BaBar} collaboration, \emph{{Measurement of the $B^0 \to \pi^\ell
  \ell^+ \nu$ and $B^+ \to \eta^{(')} \ell^+ \nu$ Branching Fractions, the $B^0
  \to \pi^- \ell^+ \nu$ and $B^+ \to \eta \ell^+ \nu$ Form-Factor Shapes, and
  Determination of $|V_{ub}|$}},
  \href{https://doi.org/10.1103/PhysRevD.83.052011}{\emph{Phys.Rev.} {\bfseries
  D83} (2011) 052011} [\href{https://arxiv.org/abs/1010.0987}{{\ttfamily
  1010.0987}}].

\bibitem{Beleno:2017cao}
{\scshape Belle} collaboration, \emph{{Measurement of the decays
  $\boldsymbol{B\to\eta\ell\nu_\ell}$ and
  $\boldsymbol{B\to\eta^\prime\ell\nu_\ell}$ in fully reconstructed events at
  Belle}}, \href{https://doi.org/10.1103/PhysRevD.96.091102}{\emph{Phys. Rev.
  D} {\bfseries 96} (2017) 091102}
  [\href{https://arxiv.org/abs/1703.10216}{{\ttfamily 1703.10216}}].

\bibitem{Ball:2007hb}
P.~Ball and G.~Jones, \emph{{$B \to \eta^{(\prime)}$ Form Factors in QCD}},
  \href{https://doi.org/10.1088/1126-6708/2007/08/025}{\emph{JHEP} {\bfseries
  0708} (2007) 025} [\href{https://arxiv.org/abs/0706.3628}{{\ttfamily
  0706.3628}}].

\bibitem{CKMfitter}
{\scshape CKMfitter Group} collaboration, \emph{{CP violation and the CKM
  matrix: Assessing the impact of the asymmetric $B$ factories}},
  \href{https://doi.org/10.1140/epjc/s2005-02169-1}{\emph{Eur. Phys. J.}
  {\bfseries C41} (2005) 1}
  [\href{https://arxiv.org/abs/hep-ph/0406184}{{\ttfamily hep-ph/0406184}}].

\bibitem{UTfit}
{\scshape UTfit} collaboration{\emph{Phys. Lett.} {\bfseries B687} (2010) 61}
  [\href{https://arxiv.org/abs/0908.3470}{{\ttfamily 0908.3470}}].

\bibitem{Kou:2018nap}
{\scshape Belle-II} collaboration, \emph{{The Belle II Physics Book}},
  \href{https://doi.org/10.1093/ptep/ptz106}{\emph{PTEP} {\bfseries 2019}
  (2019) 123C01} [\href{https://arxiv.org/abs/1808.10567}{{\ttfamily
  1808.10567}}].

\end{thebibliography}\endgroup

%



\end{document}